\documentclass[conference]{IEEEtran}
\IEEEoverridecommandlockouts

\usepackage{cite}
\usepackage{amsmath,amssymb,amsfonts}
\usepackage[linesnumbered,ruled,vlined]{algorithm2e}
\usepackage{graphicx}
\usepackage{textcomp}
\usepackage{xcolor}
\usepackage{booktabs}
\usepackage{multirow}
\usepackage{algpseudocode}
\usepackage{hyperref}  

\def\BibTeX{{\rm B\kern-.05em{\sc i\kern-.025em b}\kern-.08em
    T\kern-.1667em\lower.7ex\hbox{E}\kern-.125emX}}
\begin{document}

\SetCommentSty{mycommfont}
\newcommand{\mycommfont}[1]{\textcolor{green!50!black}{\texttt{#1}}}

\title{TurboBias: Universal ASR Context-Biasing \\ 
powered by GPU-accelerated Phrase-Boosting Tree
}

\author{
\IEEEauthorblockN{
Andrei Andrusenko\IEEEauthorrefmark{1}\IEEEauthorrefmark{3}, 
Vladimir Bataev\IEEEauthorrefmark{1}, 
Lilit Grigoryan\IEEEauthorrefmark{1},
Vitaly Lavrukhin\IEEEauthorrefmark{2}
and Boris Ginsburg\IEEEauthorrefmark{2}}
\newline
\IEEEauthorblockA{\IEEEauthorrefmark{1}\textit{NVIDIA}, Yerevan, Armenia}
\IEEEauthorblockA{\IEEEauthorrefmark{2}\textit{NVIDIA}, Santa Clara, USA}
\IEEEauthorblockA{\IEEEauthorrefmark{3}Corresponding author, \textit{aandrusenko@nvidia.com}}
}



\maketitle

\begin{abstract}
Recognizing specific key phrases is an essential task for contextualized Automatic Speech Recognition (ASR). However, most existing context-biasing approaches have limitations associated with the necessity of additional model training, significantly slow down the decoding process, or constrain the choice of the ASR system type. This paper proposes a universal ASR context-biasing framework that supports all major types: CTC, Transducers, and Attention Encoder-Decoder models. The framework is based on a GPU-accelerated word boosting tree, which enables it to be used in shallow fusion mode for greedy and beam search decoding without noticeable speed degradation, even with a vast number of key phrases (up to 20K items). The obtained results showed high efficiency of the proposed method, surpassing the considered open-source context-biasing approaches in accuracy and decoding speed. Our context-biasing framework is open-sourced as a part of the NeMo toolkit.  
\end{abstract}

\begin{IEEEkeywords}
automatic speech recognition, context-biasing, phrase-boosting, greedy decoding
\end{IEEEkeywords}

\section{Introduction}

Modern end-to-end automatic speech recognition (ASR) systems, such as Connectionist Temporal Classification (CTC)~\cite{graves2006ctc}, Recurrent Neural Transducer (RNN-T)~\cite{graves2012rnnt}, and Attention Encoder-Decoder (AED)~\cite{chan_las}, already achieve relatively high speech recognition accuracy in common data domains~\cite{Prabhavalkar2023EndtoEndSR}. However, such models often have problems recognizing specific words/phrases that are rare or absent from the training dataset (contact names, product titles, technical terms, and so on). Context-biasing methods are used to solve this problem.

The key point of context-biasing methods is the use of additional data specific to the target domain. This can be a text corpus consisting of sentences or a list of keywords/phrases whose recognition accuracy needs to be improved. In this paper, the second variant of context-biasing methods based on phrase-boosting is considered, as it is easier to prepare a context list of key phrases that may appear during the system's operation in a real-life ASR usage scenario. While collecting full-fledged text data can be difficult.

Context-biasing methods can enhance the accuracy of key phrase recognition, but this process imposes additional limitations. For example, deep-fusion methods imply introducing contextual information into the ASR model. This process requires retraining the ASR model (cross-attention approaches) \cite{Pundak2018DeepCE,Jain2020ContextualRF} or training an additional context module \cite{Yang2023PromptASRFC,Le2021ContextualizedSE,Harding2023SelectiveBW}. SpeechLM models also support the introduction of contextual information as an extra prompt \cite{Wang2023SLMBT,Chen2023SALMSL}, but this also requires using in-context learning during the model training.

Shallow fusion methods allow for the avoidance of additional model training~\cite{Zhao2019ShallowFusionEC,Jung2021SpellMN}. In this case, context-biasing is applied at the decoding stage, increasing the probability of recognizing key phrases from the context list integrated into the auxiliary boosting tree or graph. In ~\cite{Huang2024ImprovingNB}, it was shown that shallow fusion is only slightly behind deep fusion in the accuracy of keyword recognition, while maintaining the flexibility of use. 

The disadvantage of shallow fusion methods is the high slowdown of the decoding process, which must be carried out in beam search mode to expand the hypothesis search space. This problem is especially acute when decoding RNN-T and AED models, as the number of calls to the decoder module during beam search increases significantly compared to greedy mode. The solution to this problem is especially relevant for RNN-T, as this model is often chosen as a trade-off between accuracy performance, internal language model (LM) capabilities, and streaming support~\cite{He2018StreamingES,Li2020OnTC}. 

To speed up the context-biasing process, one can use a CTC-based word spotter~\cite{Andrusenko2024FastCF}, combining the results of greedy decoding from a CTC or RNN-T model with detected keywords. However, this approach requires a CTC model and only applies to offline decoding. Another way to speed up shallow fusion for RNN-T model can be the use of a GPU/TPU-friendly implementation of the Knuth-Morris-Pratt (KMP) pattern matching algorithm~\cite{Wang2023ContextualBW}. The disadvantage of the presented approach is its limited applicability to other ASR models and the lack of an open-sourced implementation.

Recent advances in accelerating RNN-T decoding using a label-looping algorithm~\cite{bataev2024labellooping} with CUDA graph implementation~\cite{galvez24_speedoflight} have reduced the difference between greedy and beam search decoding to a minimal gap, approaching the speed of greedy CTC decoding. This allowed the use of shallow fusion with a statistical n-gram language model implemented entirely on GPU (NGPU-LM)~\cite{Bataev2025NGPULMGN} in both greedy and beam search~\cite{Grigoryan2025Pushing} decoding modes without significant slowdown. The method benefits the standard context-biasing task for a specific data domain, but n-gram LM performs poorly in boosting specific isolated words. Moreover, n-gram LM requires a training text data corpus from the target domain.

This work presents GPU-PB, a universal context-biasing framework for phrase-boosting that supports all major ASR model types: CTC, Transducers (RNN-T), and AED. This framework is based on a GPU-accelerated phrase-boosting tree with a modified scoring weight distribution. This key innovation enables high-accuracy performance in both beam search and greedy decoding modes while incurring minimal computational overhead (2-5\% RTFx). The obtained results showed the superior efficiency of the proposed method, even with a massive number of key phrases (up to 20K), surpassing the considered open-source context-biasing approaches in terms of accuracy and decoding speed for speech recognition.

The main contribution of our work is the following:
\begin{itemize}
    \item \textbf{A new context-biasing framework} for fast phrase-boosting, efficient in parallel computing environments.
    \item \textbf{Modified weight distribution in boosting tree} for high-accuracy shallow fusion in greedy and beam search.
    \item \textbf{Support for all general types of ASR models}: CTC, Transducers (RNN-T), and AED.
\end{itemize}

Our context-biasing framework is openly available in the NeMo toolkit\footnote{\scriptsize{\url{https://github.com/NVIDIA/NeMo/pull/14108}}},\footnote{\scriptsize{\url{https://github.com/NVIDIA/NeMo/pull/14277}}}.

\section{Methods}

\subsection{Shallow fusion background}

The standard context-biasing process in the shallow fusion mode can be represented as follows: 

\begin{equation}
    W = \arg \max_{W} \log P(W|X) + \lambda \log P_{C}(W)
\end{equation}
Here, $W$ is the recognition result (words or tokens) of the ASR model based on the input acoustic features $X$. $P_{C}$ represents a context-biasing model applied with $\lambda$ weight. The $P_{C}$ model can be a statistical n-gram LM, neural LM, or phrase-boosting tree/graph. 

The Aho–Corasick algorithm~\cite{achocorasick}, developed for the string-searching task, can be used to build a phrase-boosting tree for this context-biasing task. Such a tree is a prefix tree constructed based on the tokenization of phrases from the context list. Each tree node is a token with the corresponding score. Failure links (or backoff in terms of n-gram LM) allow subtracting the accumulated score in the case of the absence of the required transition by token in the current tree node.

An example of an open-sourced Python implementation of phrase-boosting tree based on the Aho–Corasick algorithm is presented in the Icefall toolkit\footnote{\scriptsize{\url{https://github.com/k2-fsa/icefall/blob/master/icefall/context\_graph.py}}}. The implementation is intended for shallow fusion in RNN-T beam search decoding. However, such an implementation limits the number of requests, allowing only one token and one state per query, significantly affecting decoding speed. For efficient operation, it is necessary to obtain the probability distribution over the entire vocabulary (for example, 1024 for our models with BPE tokenization~\cite{bpe}) per request. Moreover, the basic implementation is limited to use only on the CPU.

The previously proposed NGPU-LM~\cite{Bataev2025NGPULMGN} allows you to perform queries over the entire vocabulary due to the efficient structure of storing and traversing the tree. Moreover, this work features a custom Triton kernel, which enables it to operate in shallow-fusion mode with CUDA graphs for greedy and beam-search decoding. However, this structure is designed to utilize an n-gram LM, which differs from the phrase-boosting task. Even if we build a language model only for the phrases from the context list, the tree will contain many extraneous transitions (for example, all possible unigrams as a valid start of a phrase) and their backoffs. 
This structural limitation ultimately leads to decreased accuracy in key phrase recognition, especially when the biasing list is extensive.

\subsection{Proposed approach}

\begin{figure}[t!]
  \centering
  \includegraphics[width=8.7cm]{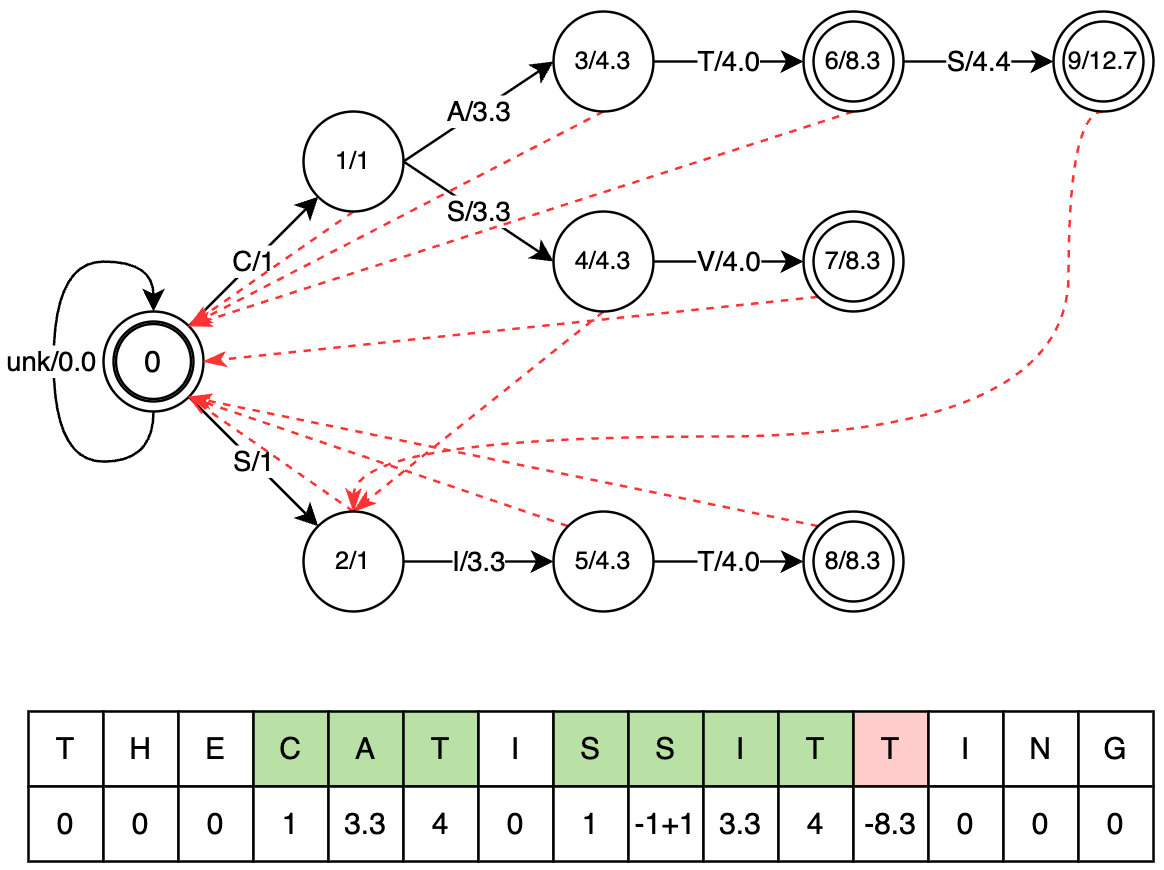}
  \caption{An example of a boosting tree for words ``CAT, CATS, CSV, SIT'' with character-level tokenization. Below the tree is an example of how boosting scores are assigned when decoding the "the cat is sitting" sequence.}
  \label{fig:pb-tree}
\end{figure}

To address the speed issue of phrase-boosting via shallow fusion and invalid transitions in NGPU-LM, we propose a new GPU-PB method that combines the efficient GPU tree architecture of NGPU-LM with the phrase-boosting tree, utilizing a modified weight distribution. 

At the first step of GPU-PB compilation, we build a standard prefix tree based on the Aho–Corasick algorithm for tokenized phrases from the context list. However, a uniform distribution of weights in the tree with additional reward (accumulated score of all transitions) only at the final node (as in the Icefall implementation) can result in low efficiency in the case of greedy search. In this case, there is no possibility of supporting multiple alternative hypotheses before reaching the final phrase node in the tree. At each moment, greedy search chooses only one hypothesis. The score value should be increased based on the depth of the tree to ensure that the boosting tree can be traversed when there is one hypothesis per decoding step. 

For more efficient score distribution, we significantly increase scores of all tree transitions with depth greater than one by the logarithm dependency $c_{0}*\beta+\ln\!\bigl(depth\bigr)$. Here, $c_{0}$ is the default context score (e.g., 1), $\beta$ is the depth scaling parameter (e.g., 2). This approach enables us to approximate the weight distribution in n-gram LM (for positive scores) and apply it to both greedy and beam search decoding.

Next, we convert the obtained prefix tree into an efficient tree structure based on NGPU-LM. For the root node, a self-loop with a score for the unknown token is added ($c_{unk}=0$ by default), intended for all tokens that are not valid beginnings of phrases from the context list (they do not have transitions from the root of the initial prefix tree). All other prefix tree arcs are converted into a new GPU-based tree, which supports sorting by the tuple ``$\mathrm{(from\_state, token)}$''. Each node also has $\mathrm{start\_arcs}$ and $\mathrm{end\_arcs}$, which ultimately allows using ``index-select'' operation for efficient score retrieval.

We save all backoff paths for each node, calculating the scores of this transition as the subtraction of the accumulated score of the current node from the score of the node to which the backoff leads. This allows subtraction of the partial score obtained from incorrect partial boosting. We also set backoffs with a zero score for each final node (end of phrase) in the tree to prevent extra score subtraction. An example of a prefix tree for context words is shown in Figure~\ref{fig:pb-tree}.

The obtained GPU-based structure enables the use of queries across the entire vocabulary through index-select operations during inference. The additional custom Triton kernel further accelerates the tree operation, enabling it to be used with CUDA Graph for CTC and RNN-T decoding. Moreover, the GPU structure allows the use of many context phrases with minimal decoding speed overhead. This can also be useful for adding different phrase forms (for example, plural) and capitalization (modern models support it by default). All this significantly increases the number of boosting phrases for the tree. The whole GPU-PB pipeline is presented in Algorithm~\ref{alg:gpu_pb}.

\begin{algorithm}[t]
\footnotesize
\caption{Three-stage algorithm for \\ phrase-boosting by the proposed GPU-PB method}
\label{alg:gpu_pb}

\SetKwInOut{Input}{Input}
\SetKwInOut{Output}{Output}
\Input{Context list $\mathcal L$ with tokenised phrases; context score $c_{0}$, depth scaling $\beta$; unknown-token score $c_{unk}$ ($0$ by default); shallow fusion weight $\lambda$.}

\vspace{0.5em}
\textbf{STAGE 1} \footnotesize{\tcp{build a prefix tree $PT$ based on the Aho-Corasick algorithm with logarithmic depth bonus}}
Initialization of the root node $n_{root}$\;
\For{phrase $\pi\in\mathcal L$}{
    $n_{cur} \gets n_{root}$ \texttt{// define current node}\;
    \For{token,depth in \textbf{enumerate}($\pi$)}{
        \eIf{$t\notin n_{cur}.next\_arcs$}{
        create a $n_{new}$ with $arc$ transition\;
        $arc.\textit{score}\gets
          \begin{cases}
            c_{0}, & depth=1\\
            c_{0}*\beta+\ln\!\bigl(d\bigr), & depth>1
          \end{cases}$\;
        \texttt{// define accumulated node score}\;
        $n_{new}.\textit{acc\_score}\gets n_{cur}.\textit{acc}+arc.\textit{score}$\;
        $n_{cur}\gets n_{new}$\;
        }
        {
        $n_{cur}\gets n_{cur}.next\_arcs[t]$\;
        }
    }
    mark $n_{cur}$ as \textit{final node}\;
}
\For{node $n$ in $PT$ with BFS order}{
      \texttt{// Add failure (backoff) links via BFS}\;
      set $\mathrm{fail}(n)$ as longest proper suffix in $PT$\;
}

\vspace{0.5em}
\textbf{STAGE 2} \footnotesize{\tcp{convert $PT$ to efficient GPU structure}}
Initialization of the table \texttt{ARCS}\;
append root self-loop $(0,\textsc{unk},\,c_{\textsc{unk}})$\;
\For{arc in $PT$ with first order}{
      append \textit{arc} parameters to \texttt{ARCS}\;
      $backoff_{score} = - n_{cur}.acc\_score$\;
}
Append self-loop transitions with $c_{unk}$ score for all tokens from the vocabulary that are not presented in $PT.root.arcs$\;

\For{arc in $PT$ with next order}{
      append \textit{arc} parameters to \texttt{ARCS}\;
      $backoff_{score} = n_{cur}.fail.acc\_score - n_{cur}.acc\_score$\;
}

sort \texttt{ARCS} by \emph{(from\_state, token)} and upload to GPU\;
\textit{BT}$\gets$PyTorch \texttt{nn.Module} Graph Wrapper

\vspace{0.5em}
\textbf{STAGE 3} \tcp{ASR model inference}
Initialization of \textit{BT} states $S$\;
\For{decoding step}{
      $\mathbf{t}_{1:B}\gets$ last tokens of current beam (or best) hypotheses $hyps$\;
      \eIf{GPU-PB enabled}{
          $BT_{scores},S_{next}\gets\textit{BT.get\_scores}(S[\mathbf{t}_{1:B}])$\;
          $\textit{hyps\_expansion\_score}\gets acoustic_{scores}+ \lambda\cdot BT_{scores}$\;
      }{
          $\textit{hyps\_expansion\_score}\gets acoustic_{scores}$\;
      }
      \texttt{// Select one or beam best hypotheses}\;
      $hyps = prune\_expanded(hyps, hyps\_expansion\_score)$\;
}
\end{algorithm}

\subsection{Adaptation for ASR models}

GPU-PB has a similar structure to NGPU-LM. The proposed phrase-boosting method can also be applied to all major ASR models in greedy and beam search decoding modes, including CTC, Transducers, and AED, using efficient decoding pipelines with NGPU-LM~\cite{Bataev2025NGPULMGN,Grigoryan2025Pushing,article:flexCTC}.

For greedy decoding of both CTC and RNN-T models, the two-stage token selection~\cite{Bataev2025NGPULMGN} is performed on each decoding step. First, greedy selection is applied without boosting. If the result is a blank token, we proceed to the next decoding step. For CTC models, we also preserve the repeated token if selected, without applying a boosting weight to it, and proceed to the next step. Otherwise, we reweight the rest of the tokens (non-blank tokens for RNN-T, non-blank and not repeated tokens for CTC) using the boosting tree and reselect the output greedily from this category.

In AED models, the end-of-sentence (EOS) symbol is used as an auxiliary token, which controls the end of the decoding process. NGPU-LM can directly predict the final weight during decoding, since it is a complete statistical language model. In the case of GPU-PB, we have no way to get a fair EOS score from the boosting tree. In this case, the scores from the boosting tree can exceed the EOS score, which can lead to AED decoding hallucination at the end of the sentence. To solve this problem, we use the following heuristics -- at each decoding step, we increase the default EOS score by the maximum score obtained from the boosting tree. We also increase the EOS score by the value of the final weight of the node when detecting a whole phrase from the boosting tree. These techniques allow us to cope with EOS suppression in the case of AED models.

Another feature of greedy decoding with GPU-PB for all ASR models is the potential for recognition accuracy degradation of words not included in the context list. This is probably linked with the distortion of such words from the boosting tree due to the difference in scores between the unknown token and the first transition in the boosting tree, limited by the absence of several parallel recognition hypotheses. In this case, we recommend using an unknown score value close to the score of the first transition along the tree $c_{0}$.


\section{Experimental setup}

\subsection{ASR models}

For the method evaluation, we used ASR models trained on a large amount of data to get closer to a real-world scenario. To achieve this, we utilized a 24,000-hour set of open data with labeled English speech and a BPE tokenizer \cite{bpe} with 1024 tokens.

As ASR models, we trained CTC, RNN-T, and AED (Canary~\cite{canary}) models with the same FastConformer encoder architecture~\cite{rekesh2023fastconformer} with 108M parameters. 
For the convenience of training and general use of the CTC and RNN-T models, we trained a Hybrid Transducer-CTC model \cite{noroozi2024stateful} (a shared encoder trained together with CTC and Transducer output heads). 
For the RNN-T head, a single-layer LSTM prediction network (decoder) of size 640 was used, which increased the total model size to 114M parameters.
For the AED model, we utilized a 4-layer Transformer (FF 1024, Self-attention 256) as the decoder, resulting in a final model size of 179M.

Model evaluations were performed in greedy and beam search decoding with a beam size of 8, except for the AED model. For AED, we used a beam size of 3, as increasing the beam further resulted in a significant decoding slowdown. 

\subsection{Evaluation data}

We used several test datasets specific to a particular domain to evaluate the context-biasing task.

\textbf{Earnings21}. Earning21~\cite{delrio2021earnings21} represents about 40 hours of earnings calls, typical for the financial data domain. The dataset consists of long records, up to 1.5 hours in length, which can cause recognition problems for many ASR models. We segmented the dataset into samples up to 30 seconds. As a dev and test sets, we selected subsets of 5 and 10 hours, respectively. For the context list, we used the baseline oracle list\footnote{\scriptsize{\url{https://github.com/revdotcom/speech-datasets/blob/main/earnings21/bias_lists/oracle_list.txt}}}, removing short phrases of 2 letters in length to prevent an increase in the false accept rate for such cases. As a result, we obtained a context list with 893 phrases.

\textbf{MultiMed}. Additionally, we prepared a subset from the MultiMed medical dataset~\cite{le2024multimed}. We further processed the test and dev sets by filtering out files with incorrect transcriptions based on the recognition results of the base model (WER $>$ 30\%). The final dataset sizes were 13.5 hours for the test and 7.3 hours for the dev. 
To select specific boosting phrases, we used the LLaMA 3.3-70B Instruct model~\cite{model:llama3_70b_instruct}, prompted to tag medical terms, such as diseases, procedures, medications, and clinical conditions. To identify named entities such as personal names, brand names, and organizations, we used the spaCy toolkit~\cite{article:spacy}. We filtered the resulting phrases based on the recognition results of the baseline ASR model, eliminating simple ($>$70\% accuracy) and low-frequency ($<$10 occurrences) phrases. We obtained a list of 991 unique phrases.

\begin{table}[t!]
  \centering
  \caption{Statistics of evaluation datasets.}
  \label{tab:dataset_stats}
  \begin{tabular}{lccc}
    \toprule
    \textbf{Metric} & \textbf{CSTalks} & \textbf{Earnings21} & \textbf{MultiMed} \\
    \midrule
    Dev set (h)            & 3.2   & 5.2  & 7.3  \\
    Test set (h)           & 8.3   & 10.0 & 13.5  \\
    Context list size         & 200 & 893 & 991 \\
    Occurrences in test     & 3500 & 560 & 4541 \\
    \midrule
    \multirow{3}{*}{Example phrases}        & \texttt{gpu, cpu} & \texttt{nokia} & \texttt{aorta} \\
     &  \texttt{pytorch} & \texttt{deloite} & \texttt{acupuncture} \\
     & \texttt{copilot} & \texttt{glenrock} & \texttt{lymphatics} \\
    \bottomrule
  \end{tabular}
\end{table}

\textbf{CSTalks}. We collected open-keynotes from technology companies that contain many specific terms to test the proposed method in the modern computer science domain. After text processing and audio segmentation, we obtained 3.2 hours for dev and 8.3 hours for test sets. We used the same methodology as for the MultiMed datasets to prepare the phrase-boosting list. The resulting context-biasing list consists of 200 unique words with high occurrence density in the sets.

Table~\ref{tab:dataset_stats} provides general information on the evaluation data used with a detailed specification. The dev sets were used for boosting parameters' search, and the test for obtaining the final results in Section~\ref{sec:results}.

\subsection{Metrics}

We used the F-score ($2*Precision*Recall/(Precision+Recall)*100\%$) to measure key phrase recognition accuracy, which was calculated for phrases from the context list. Additionally, we calculated the overall WER value for the entire recognition result to assess the overall recognition quality of the ASR systems. To track the decoding speed, we measured the inverse Real-Time Factor (RTFx). RTFx is calculated as the total audio duration divided by the ASR model evaluation time. RTFx is measured after a single warm-up run as the average value of 3 runs on a single NVIDIA RTX A6000 GPU with a batch size of 32.

\begin{table*}[t]
  \centering
  \caption{Performance evaluation of the proposed GPU-PB method in the context-biasing task for CTC, RNN-T, and AED models in greedy and beam search decoding modes. The GPU-PB column is responsible for enabling phrase-boosting. Precision and recall are rounded for the convenience of perception. F-score and WER are in \%; RTFx is inverse real-time factor averaged across all sets.}
  \label{tab:overalll_results}
  \resizebox{\textwidth}{!}{%
  \begin{tabular}{llc|cc|cc|cc|c}
    \toprule
    \multirow{2}{*}{\textbf{Model}} & \multirow{2}{*}{\textbf{Decod}} & \textbf{GPU} &
      \multicolumn{2}{c}{\textbf{CSTalks}} &
      \multicolumn{2}{c}{\textbf{Earnings21 (10h)}} &
      \multicolumn{2}{c|}{\textbf{MultiMed}} & \textbf{RTFx}$\uparrow$ \\
    \cmidrule(lr){4-5}\cmidrule(lr){6-7}\cmidrule(lr){8-9}
      & & \textbf{PB} & \textbf{F-score (P/R)}$\uparrow$ & \textbf{WER}$\downarrow$ &
        \textbf{F-score (P/R)}$\uparrow$ & \textbf{WER}$\downarrow$ &
        \textbf{F-score (P/R)}$\uparrow$ & \textbf{WER}$\downarrow$ & \textbf{Avg.} \\
    \midrule
    \multirow{4}{*}{CTC}
      & \multirow{2}{*}{greedy}   & --         & 35.0\quad(97/21) & 13.7 & 45.7\quad(94/30) & 15.6 & 54.0\quad(95/38) & 15.0 & 2181 \\
      &                           & \checkmark & 64.8\quad(94/50) & 11.9 & 53.5\quad(92/38) & 15.6 & 60.2\quad(93/45) & 14.9 & 2067 \\
      \cmidrule(lr){2-10}
      & \multirow{2}{*}{beam}     & --         & 35.0\quad(97/21) & 13.8 & 45.7\quad(94/30) & 15.6 & 54.0\quad(95/38) & 15.0 & 1874 \\
      &                           & \checkmark & \textbf{83.2}\quad(90/77) & \textbf{10.2} & \textbf{67.8}\quad(89/55) & \textbf{15.5} & \textbf{71.8}\quad(89/60) & \textbf{14.3} & 1786 \\
    \midrule
    \multirow{4}{*}{RNN-T}
      & \multirow{2}{*}{greedy}   & --         & 42.5\quad(96/27) & 12.8 & 56.0\quad(95/40) & 15.1 & 60.4\quad(95/44) & 13.9 & 1822 \\
      &                           & \checkmark & 70.4\quad(92/57) & 10.7 & 63.3\quad(93/48) & 15.0 & 66.3\quad(91/52) & 13.7 & 1751 \\
      \cmidrule(lr){2-10}
      & \multirow{2}{*}{beam}     & --         & 44.2\quad(97/29) & 12.8 & 55.8\quad(93/40) & 14.3 & 62.5\quad(95/47) & 13.6 & 1466 \\
      &                           & \checkmark & \textbf{82.9}\quad(90/76) &  \textbf{9.6} & \textbf{74.0}\quad(88/64) & \textbf{14.2} & \textbf{75.8}\quad(89/66) & \textbf{12.9} & 1420 \\
    \midrule
    \multirow{4}{*}{AED}
      & \multirow{2}{*}{greedy}   & --         & 52.6\quad(97/36) & 12.7 & 54.7\quad(92/39) & 15.4 & 64.0\quad(94/49) & 14.1 &  356 \\
      &                           & \checkmark & 75.6\quad(93/64) & 10.4 & 63.8\quad(91/49) & 15.3 & 69.3\quad(89/57) & 13.9 &  350 \\
      \cmidrule(lr){2-10}
      & \multirow{2}{*}{beam}     & --         & 53.7\quad(97/37) & 12.5 & 54.6\quad(92/39) & 15.2 & 65.7\quad(94/51) & 13.7 &  145 \\
      &                           & \checkmark & \textbf{82.4}\quad(94/73) & \textbf{10.2} & \textbf{66.2}\quad(88/53) & \textbf{15.1} & \textbf{75.5}\quad(88/66) & \textbf{13.1} &  141 \\
    \bottomrule
  \end{tabular}%
  }
\end{table*}

\subsection{Comparison with other methods}

To compare the proposed method, we investigated other open-source solutions with phrase-boosting support: Pyctcdecode\footnote{\scriptsize{\url{https://github.com/kensho-technologies/pyctcdecode}}}, CTC-WS~\cite{Andrusenko2024FastCF}, and NGPU-LM~\cite{Bataev2025NGPULMGN}. Pyctcdecode is a CTC decoding framework that has beam search decoding with word boosting in shallow fusion mode.
CTC-WS is a method that combines CTC/RNN-T greedy decoding results with key phrases detected by a word spotter. Word spotter uses CTC-logits to search for words in the fast beam search decoding mode over the context-biasing graph. The need for CTC-logits requires the presence of a hybrid Transducer-CTC architecture to work with RNN-T. For NGPU-LM method, we built a 6-gram token-level language model based on the context list.

We also investigated GPU-PB with uniform weight distribution and additional hypothesis reward in the final tree node (GPU-PB\textsubscript{uw}), similar to the baseline implementation of the boosting tree in the Icefall framework.

\section{Results}
\label{sec:results}

For all the presented results on the test sets, we experimentally selected the boosting weight parameters for each method in such a way as to obtain the minimum possible WER value on development sets.

\subsection{Proposed phrase-boosting method}

The overall results of applying the proposed context-biasing method are presented in Table~\ref{tab:overalll_results}. GPU-PB improved the accuracy of key phrase recognition (F-score) and also achieved additional improvements in overall speech recognition accuracy (WER) for almost all cases considered. This effect is most noticeable for the CSTalks dataset, as it has the highest density of key phrases from the context list, for which recognition accuracy can be improved.

\textbf{Greedy vs Beam}. One of the important features of GPU-PB is the demonstrated ability to improve the results in greedy decoding mode (8-10\% absolute F-score on average), since this allows boosting phrases in streaming decoding. To get the best performance from GPU-PB, beam search is necessary (17-23\% absolute F-score on average), since processing several hypotheses simultaneously allows expanding the search space of key phrases from the context list. However, current versions of beam search apply only to offline decoding scenarios and are 11-20\% slower than greedy decoding. 

\textbf{RTFx}. GPU-PB demonstrated an average slowdown of 2-5\% in RTFx across all ASR models and datasets considered (including test sets with a context list size of up to 1000 phrases -- MultiMed), even in the case of beam search. This fact proves the high efficiency of using the GPU implementation of phrase-boosting for greedy and beam decoding.


\textbf{ASR models}. The most balanced choice in terms of accuracy and decoding speed for the context-biasing task was the RNN-T model, due to the presence of a decoder block and the effective implementation of beam search using CUDA graphs. GPU-PB also showed high efficiency for the AED model in greedy and beam search decoding. However, the speed of such an AED model is significantly lower than CTC and RNN-T. Nevertheless, these results open up prospects for utilizing GPU-PB with SpeechLM models that employ an autoregressive Transformer decoder.

\subsection{Comparison with other methods}

The results of the comparison with other existing open-source solutions are presented in Table~\ref{tab:cb_comparison}. We evaluated all the methods on the CSTalks dataset using both CTC and RNN-T models. The proposed GPU-PB method demonstrated the best efficiency in the case of beam search, outperforming the considered methods by 2-18\% in absolute F-score and exhibiting high decoding speed.


The CTC-WS approach showed the best accuracy in greedy decoding due to the presence of an additional word detector, which requires access to the full audio file. This limitation restricts the use of CTC-WS in streaming mode. Moreover, CTC-WS is slower than GPU-PB by about 2 times due to the word detector running on the CPU.

Uniform weight distribution for GPU-PB (GPU-PB\textsubscript{uw}) yielded good results. However, the accuracy of key phrase recognition for this method is limited due to the lengthy process of weight accumulation for the hypothesis, which is especially important in greedy mode.

Pyctcdecode also showed an improvement relative to the baseline with almost the same accuracy as GPU-PB\textsubscript{uw} for CTC. However, the speed of pyctcdecode degrades significantly with an increase in the number of boosting phrases due to the suboptimal shallow fusion implementation. 

Using NGPU-LM also improves accuracy, but the presence of a large number of invalid transitions in the n-gram language model limits the effectiveness of this method. This fact is especially acute in beam decoding, where the results for this method are even worse than in greedy mode.


\begin{table}[t!]
  \centering
  \caption{Performance comparison of considered context-biasing strategies for the phrase-boosting task. F-score shows precision/recall; WER in \%; RTFx is inverse real-time factor.}
  \label{tab:cb_comparison}
  \resizebox{\columnwidth}{!}{%
  \begin{tabular}{ll|cc|c}
    \toprule
    \textbf{Decod.} & \textbf{C-Biasing} & \textbf{F-score (P/R)}$\uparrow$ & \textbf{WER}$\downarrow$ & \textbf{RTFx}$\uparrow$ \\
    \midrule
    \textbf{CTC} \\
    \midrule
    \multirow{5}{*}{greedy}
      & --              & 35.0\quad(97/21) & 13.7 & 2232 \\
      & CTC-WS          & \textbf{79.8}\quad(90/72) & \textbf{10.9} &  906 \\
      & NGPU-LM          & 58.5\quad(96/42) & 11.9 & 2123 \\
      & GPU-PB\textsubscript{uw}      & 53.7\quad(96/37) & 13.2 & 2181 \\
      & GPU-PB          & 64.8\quad(94/50) & 11.9 & 1991 \\
      \midrule
    \multirow{5}{*}{beam}
      & --              & 35.0\quad(97/21) & 13.8 & 1883 \\
      & Pyctcdecode     & 74.0\quad(93/62) & 11.5 &   29 \\
      & NGPU-LM          & 55.2\quad(97/39) & 12.5 & 1807 \\
      & GPU-PB\textsubscript{uw}      & 75.0\quad(94/62) & 11.5 & 1784 \\
      & GPU-PB          & \textbf{83.2}\quad(90/77) & \textbf{10.2} & \textbf{1777} \\
    \midrule
    \textbf{RNN-T} \\
    \midrule
    \multirow{5}{*}{greedy}
      & --              & 42.5\quad(96/27) & 12.8 & 1832 \\
      & CTC-WS          & \textbf{80.0}\quad(90/72) & \textbf{10.1} &  639 \\
      & NGPU-LM          & 68.5\quad(96/54) & 10.8 & 1812 \\
      & GPU-PB\textsubscript{uw}      & 65.0\quad(94/50) & 12.1 & 1759 \\
      & GPU-PB          & 70.4\quad(92/57) & 10.7 & 1753 \\
      \midrule
    \multirow{4}{*}{beam}
      & --              & 44.2\quad(97/29) & 12.8 & 1467 \\
      & NGPU-LM          & 58.6\quad(97/42) & 12.0 & 1426 \\
      & GPU-PB\textsubscript{uw}      & 80.9\quad(93/72) & 10.1 & 1417 \\
      & GPU-PB          & \textbf{82.9}\quad(90/76) & \textbf{9.6} & \textbf{1430} \\
    \bottomrule
  \end{tabular}%
  }
\end{table}

\subsection{Robustness}

To evaluate the robustness of the proposed GPU-PB method, we investigated its performance by increasing the context-biasing list to 20K phrases. For this experiment, we used the CSTalks dataset and the RNN-T model in beam search decoding mode. We left the beam size at 8 and slightly reduced the boosting weight from 1 to 0.7 to avoid a large number of false accepts during decoding. The baseline context list size for CSTalks is 200 phrases. Next, we generated approximately 20K special terms, each 1 to 3 words in length, using the LLM model, and gradually expanded the base context list to include up to 20K phrases.

\begin{figure}[t!]
  \centering
  \includegraphics[width=\columnwidth]{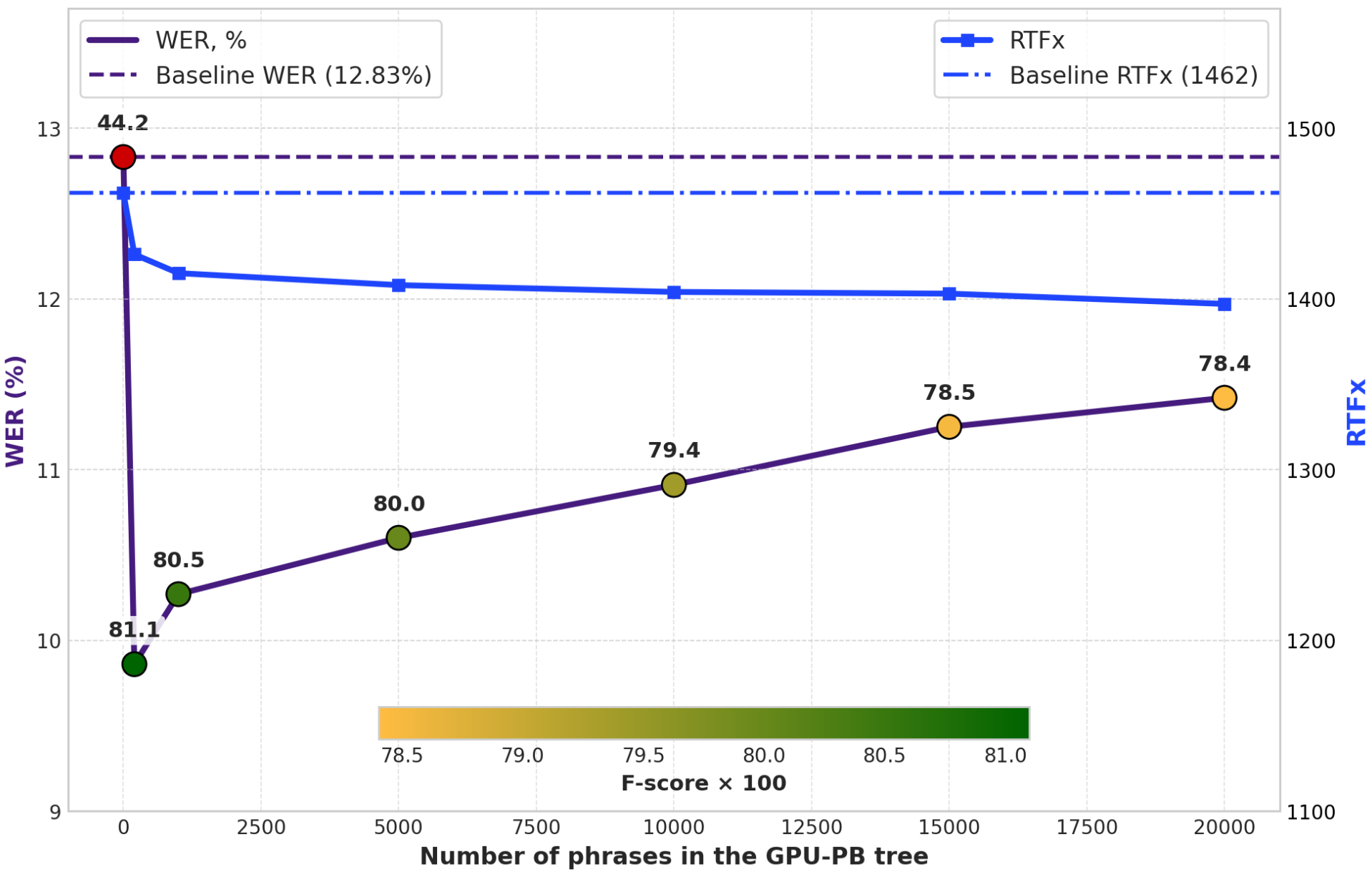}
  \caption{GPU-PB robustness to the context list size. The left y-axis represents WER (purple). The right y-axis represents RTFx (blue). The dashed lines represent the baseline WER and RTFx values without phrase-boosting. The F-score is represented by gradient circles with exact values on the WER graph. The red color is for the lowest F-score, and green is for the highest.}
  \label{fig:cb_list_robustness}
\end{figure}

The results of the described experiment are presented in Figure~\ref{fig:cb_list_robustness}. Based on the obtained results, we can conclude that the proposed method is relatively robust to variations in the context list size. Increasing the number of boosting phrases from 0 to 20,000 slowed down the entire system runtime by only about 5\% (blue plot). Recognition accuracy has deteriorated from 81.1\% to 78.4\% F-score (gradient circles) and from 9.86\% to 11.42\% WER (purple plot) due to additional false accepts, but remains noticeably better than the baseline (44.2\% F-score and 12.83\% WER). 

\subsection{Complementarity}

We also explored the possibility of using GPU-PB together with NGPU-LM for MultiMed data in CTC beam search decoding. In this case, phrases from the context list were used for phrase-boosting. For LM fusion, we built an n-gram LM based on the training text data. This combination helped to reduce the WER on the test set from 15.09\% to 13.55\%, while using the methods alone showed 14.47\% and 14.00\% WER, respectively. The results demonstrated the complementarity of the considered approaches, thereby improving the overall recognition result.

\section{Conclusion}

In this paper, we introduced a universal ASR context-biasing framework that supports all major ASR models, including CTC, Transducers (RNN-T), and AED models. This framework utilizes a GPU-accelerated phrase-boosting tree with a modified scoring weight distribution for efficient shallow fusion decoding. The obtained results showed an improvement of key phrases recognition (F-score) by 8-10\% in greedy and 17-23\% absolute in beam search decoding, with an additional improvement in overall WER relative to the baseline without phrase-boosting. The speed overhead of using phrase-boosting consumed only 2-5\% RTFx of the total ASR running time, demonstrating high robustness to context list size growth up to 20K items. The proposed method also surpasses the considered open-source context-biasing approaches in accuracy and decoding speed. The proposed context-biasing framework is open-sourced as a part of the NeMo toolkit. 



\bibliographystyle{IEEEtran}
\bibliography{refs}

\end{document}